# Properties of Cross-Impact Balance Analysis

*Wolfgang Weimer-Jehle[1]*

*December 29th, 2009*

**Abstract**

*CIB matrices are N x N hypermatrices, the elements of which are m x n matrices. They are used in Cross-Impact Balance Analysis, a concept applied in social sciences, management sciences, scenario analysis and technology foresight to identify plausible configurations of qualitatively defined impact networks. Cross-Impact Balance Analysis (CIB) offers an opportunity for qualitativ systems analysis without complex mathematics. Although CIB doesn't confront its user with too much mathematics, the background of the method and its algorithm can be scrutinized by mathematical means, thus revealing an extensive set of useful properties which are described and proved in this article. Among them are four laws of invariance, a treatise on several special cases of CIB matrices, and the proof that CIB analysis is equivalent to a universal computer (a Turing machine).*

## 1. Introduction

The application of mathematical simulation methods in a social science context is fraught with difficulty. Probably the most critical aspect is that essential knowledge about economical, technological, social, and political systems and their interdependence is often restricted to qualitative insights and implicit mental models produced by experts. On the other hand, there is clearly a need for a suitable mathematical approach, because the state of extreme interdependence inside and between these systems frequently excludes the possibility of intuitive systemic understanding.

In the past, several techniques have been developed to meet the needs of interdependence analysis under the special conditions of multidisciplinary systems that include social systems. One of the most popular is a group of techniques denoted as *Cross-Impact Analysis*, originally introduced by Gordon and Hayward in 1968 [1]. In these techniques, expert judgements about the interdependence of the main system variables are collected in a matrix scheme, and a more or less heuristic evaluation procedure is used to compute scenarios of probable system behaviour.

Cross-Impact Analysis achieved considerable popularity among those concerned with projecting and analysing scenarios to do with political, economical, technological, or social change. However, the method has also been the focus of criticism. Part of the criticism was that expert judgements should not be used as a base for complex quantitative calculations. In this respect the recently proposed cross-impact method, Cross-Impact Balance Analysis (CIB [3]) promises progress. It combines a qualitative orientation and a lucid algorithm with considerable analytic power. To achieve this it introduces a new type of matrix calculus which requires no advanced mathematical education of the user. However, a mathematical viewpoint helps to find many useful properties of this matrix type. In the following chapers a short introduction to CIB, some basic definitions and a set of properties and their proofs of CIB matrices are given.

[1] University of Stuttgart, Institute for Social Sciences V, Interdisciplinary Research Unit on Risk Governance and Sustainable Technology Development (ZIRN), Seidenstr. 36, 70174 Stuttgart, Germany. Phone +49 711 685 84301. Fax +49 711 685 82487. Email: wolfgang.weimer-jehle@sowi.uni-stuttgart.de

## 2. CIB basics

The Cross-Impact Balance Analysis (CIB Analysis) was developed in 2001 to overcome several practical difficulties inherent to established cross-impact methods [2]. The method was described in detail by Ref. [3] and a software tool developed to support the application of the method [4]. CIB was applied and developed in various scenario projects dealing with political/technological/environmental interactions in the energy sector [5]-[8], with innovation systems [9], [10], with global carbon emission scenarios [11], and with syndrome phenomena in sustainability [12], [13]. CIB's concept gets its inspiration from systems theory and finite automata theory [3], [14].

CIB offers a simple tool to evaluate qualitative expert insights in complex multidisciplinary systems in order to construct consistent scenarios of the system state. Experts (recognized authorities in the field) are asked if the occurrence of a certain state of a system variable ("descriptor") will promote or restrict a certain state of an other system variable. The expert judgements on all pair interactions within a system are collected in the "cross-impact matrix". The assessment of the factor relationships is usually carried out on a integer scale with positive values for promoting direct impacts and negative values for restricting direct impacts. Frequently a [-3...+3] scale is sufficient, but higher values may also be used to express particularly strong impacts.

Every descriptor state can in principle be influenced by all other descriptors. The cross-impact matrix is a database that answers the question as to how a certain state impacts a certain other state. If a state is the target of several impacts, the superposition of impacts must be taken into account. The combined effect of several impacts is modelled in CIB by adding up the appropriate cross-impact judgements. In this way, the internal impact flows of a given system state (a "scenario") can be visualized in a CIB matrix by highlighting the rows of the corresponding states and summing them up to produce *impact balances*.

Table 1 shows a simple example. It deals with the opinion formation within a group of persons with respect to an arbitrary question of public concern. In the Tab. 1 example the group members are partly acquainted with each other, and in general the opinion of a person is influenced by the opinions of his/her acquaintances. In Table 1, Tom is not acquainted with Nancy and therefore there is no *direct* influence between these persons ($C_{12}$ and $C_{21}$ are zero). Tom and Helen are acquainted with each other. Tom thinks highly of Helen and strongly rates her opinion. The opposite is not the case, unfortunately ($C_{41}$ contains strong cross-impact values, $C_{14}$ is zero). Ray is strongly motivated to adopt the opposite opinion to Max but tends to be indifferent if Max offers no opinion. Nancy notes Helen's opinion but only in instances where Helen agrees with a statement. Max listens to Tom and Nancy, but Nancy's opinion usually carries more weight in cases of disagreement. Diagonal judgement sections are zero for regular CIB matrices.

These relationships (and various others shown in Tab. 1) constitute a net of impact relations expressed by the cross-impact matrix. They limit the space of plausible scenarios for the system state because an opinion configuration chosen arbitrarily will in general contain contradictions to the "rules" of the group. Contradictions are made visible by calculating the impact balances of a scenario. In Table 1 this is made for the scenario

$$\underline{z} = [0,-,+,+,-]$$

(Tom is indifferent, Nancy disagrees, Ray and Helen agree, Max disagrees).

Summing up the highlighted rows summarizes the several influences impacting the descriptors. The scores of the impact balances which correspond to the given scenario are marked by arrows. In three cases (Ray, Helen, Max), the arrows indicate the maximum score of the descriptor impact balance. Consider Helen as an example. She is assumed to agree and this coincides with the fact that Ray agrees (the only person who influences her).

**Tab. 1:** Social net CIB matrix: coupled opinion formation. The state "+" symbolizes a person's approval of a given statement, "-" symbolizes the rejection of the statement. The row descriptor states are impact sources, the column descriptor states are impact sinks. $C_{25}(1,3) = -2$ indicates that Nancy's agreement to a statement will moderately hinder Max from rejecting this statement.

|  | 1.Tom | | | 2.Nancy | | | 3.Ray | | | 4.Helen | | | 5.Max | | |
|---|---|---|---|---|---|---|---|---|---|---|---|---|---|---|---|
|  | + | 0 | - | + | 0 | - | + | 0 | - | + | 0 | - | + | 0 | - |
| **1.Tom:** | | | | | | | | | | | | | | | |
| + | | | | 0 | 0 | 0 | 0 | 0 | 0 | 0 | 0 | 0 | 1 | 0 | -1 |
| 0 | | | | 0 | 0 | 0 | 0 | 0 | 0 | 0 | 0 | 0 | 0 | 0 | 0 |
| - | | | | 0 | 0 | 0 | 0 | 0 | 0 | 0 | 0 | 0 | -1 | 0 | 1 |
| **2.Nancy:** | | | | | | | | | | | | | | | |
| + | 0 | 0 | 0 | | | | 0 | 0 | 0 | 0 | 0 | 0 | 2 | 0 | -2 |
| 0 | 0 | 0 | 0 | | | | 0 | 0 | 0 | 0 | 0 | 0 | 0 | 0 | 0 |
| - | 0 | 0 | 0 | | | | 0 | 0 | 0 | 0 | 0 | 0 | -2 | 0 | 2 |
| **3.Ray:** | | | | | | | | | | | | | | | |
| + | 0 | 0 | 0 | 0 | 0 | 0 | | | | 2 | 0 | -2 | 0 | 0 | 0 |
| 0 | 0 | 0 | 0 | 0 | 0 | 0 | | | | 0 | 0 | 0 | 0 | 0 | 0 |
| - | 0 | 0 | 0 | 0 | 0 | 0 | | | | -2 | 0 | 2 | 0 | 0 | 0 |
| **4.Helen:** | | | | | | | | | | | | | | | |
| + | 3 | 0 | -3 | 2 | 0 | -2 | 2 | 0 | -2 | | | | 0 | 0 | 0 |
| 0 | 0 | 0 | 0 | 0 | 0 | 0 | 0 | 0 | 0 | | | | 0 | 0 | 0 |
| - | -3 | 0 | 3 | 0 | 0 | 0 | -2 | 0 | 2 | | | | 0 | 0 | 0 |
| **5.Max:** | | | | | | | | | | | | | | | |
| + | 2 | 0 | -2 | 1 | 0 | -1 | -3 | 0 | 3 | 0 | 0 | 0 | | | |
| 0 | 0 | 0 | 0 | 0 | 0 | 0 | -1 | 2 | -1 | 0 | 0 | 0 | | | |
| - | -2 | 0 | 2 | -1 | 0 | 1 | 3 | 0 | -3 | 0 | 0 | 0 | | | |
| **Balance:** | 1 | 0 | -1 | 1 | 0 | -1 | 5 | 0 | -5 | 2 | 0 | -2 | -2 | 0 | 2 |

$C_{25}(1,3)$: A judgement cell
A judgement group
$C_{43}$: A judgement section

The impact score of a state
The impact balance of a descriptor

The judgement values in Tab. 1 sum to zero in each judgement group. This "standardization" is not required by the CIB algorithm but it enhances logical consistency, as promoting influences towards one state restricts respective opposites. The use of standardized judgement groups is therefore recommended in CIB analysis.

In two descriptor balances (Tom and Nancy) the arrows do not point to the maximum impact score and this indicates inconsistencies in the assumed scenario: Tom is assumed to be indifferent and Nancy is assumed to disagree, although Helen agrees (outvoting the influence of Max, who disagrees). Both descriptors violate the "rules" which are coded in the cross-impact matrix. To avoid such violations, the states of the descriptors must show a well-balanced configuration that reflects the dual role of each descriptor as both impact source and impact sink.

The self-consistency of a scenario $\underline{z} = [z_1, z_2, ...]$ requires that every state is chosen in such a way as to ensure that no other state of the same descriptor is preferred more strongly by the combined influences of the other descriptors. In CIB this is denoted as *the principle of consistency*. It follows that a consistent scenario $\underline{z}$ is a solution of the inequality:

$$\sum_i C_{ij}(z_i, z_j) \geq \sum_i C_{ij}(z_i, l) \qquad (1)$$

valid for all $j$, $l$. The indices $i$ and $j$ run over all descriptors, $l$ runs over all states of a descriptor. For the definition of the matrix cell $C_{ij}(k,l)$ see Table 1. The inequality (1) acts as an effective restriction to the set of possible scenarios. Actually, in the Table 1 example only 2 scenarios out of $3^5=243$ possible ("combinatorial") scenarios are solutions of (1). The two consistent scenarios are:

a) [0,0,0,0,0]

All group members are indifferent to the statement in question.

b) [-,+,-,-,+]

Nancy and Max agree. All others disagree.

The consistency of these solutions can be easily checked by applying the procedure shown in Table 1. Consistent scenarios reflect the effect of the indirect influences of a system, as can be seen in our example: The consistent scenarios exhibit a correlation of Tom's and Ray's opinion although there is no direct influential link between them.

The solutions of (1) can be found simply by enumerating all combinatorial scenarios, carrying out the consistency check shown in Table 1 for every step. But there are other search methods. One of them, CIB succession, reveals links with automata network research:

Table 1 shows the consistency check of the scenario

[0,-,+,+,-].

It demonstrates inconsistencies in the descriptors 1 and 2 (Tom and Nancy). A naive attempt to erase these inconsistencies would be to change the scenario states of the inconsistent descriptors to the states of maximum impact scores. In the example, this leads to the scenario

[+,+,+,+,-].

Unfortunately, this adjustment does not usually result in a consistent scenario because the adjustments of the inconsistent descriptors often cause new inconsistencies in other nodes of the impact network. A few repetitions of this procedure may however lead to a consistent scenario (provided that a consistent scenario exists). In the Tab. 1 example, the sequence of scenarios ("CIB succession") reads:

[0,-,+,+,-] → [+,+,+,+,-] → [+,+,+,+,+] →

[+,+,-,+,+] → [+,+,-,-,+] → [-,+,-,-,+]

The last scenario is consistent and further succession steps do not alter it: each consistent scenario is a CIB succession attractor. To find all solutions of (1), the succession procedure has to be repeated with each combinatorial scenario as a starting scenario. The succession rule can be defined in different ways. We describe four of them:

a) Adjust all inconsistent descriptors to the state of highest impact score (global succession adjustment). This is the succession rule described above. It is usually used in CIB analysis.

b) Adjust all inconsistent descriptors towards the state of highest impact score, but only by jumping to a neighbour state (incremental succession adjustment). This rule attempts to approach the dynamics of differential equations by restricting the distance between a scenario and its successor.

c) Adjust only the descriptor(s) with the highest inconsistency and change it to the state of the highest impact score (local succession adjustment). This rule expresses the idea that the reaction of the system elements is accelerated by high system forces and that the system element sensing the greatest forces reacts first.

d) Adjust only the first inconsistent descriptor to the state of highest impact score (adiabatic succession adjustment). This rule is suitable if the descriptor reaction times feature distinct time scales, and the descriptors are sorted according to time scale, starting with the fastest descriptor. In synergetics, an adiabatic dynamic is associated with the possible emergence of self-organization phenomena [15], [16].

In case of a tie there might be more than one possibilities to proceed with the succession.

All described succession types lead strictly to the same set of consistent scenarios. But they may lead to different combinatorial weights and to different cyclic attractors (cf. section 3 / IX).

It should be stressed that the application of CIB is not restricted to the problem of opinion formation, which is used here only as an example. Typical descriptors in the application projects afore-mentioned in this section concern policy decisions, business strategies, environmental, social or techological changes and others.

CIB also goes far beyond the example of Table 1 mathematically. The number of descriptor states may be chosen arbitrarily. The descriptors may show different numbers of states. The judgement scale shown in Table 1 is not limited to the range [-3...+3]: Higher numbers can be used to express stronger influences. In addition, CIB calculus is not restricted to integer numbers, although experts will usually prefer to make their judgements in this format. From a mathematical point of view, every mathematical object is suitable to be a cross-impact judgement in CIB framework, provided that an addition and a greater/smaller comparison can be defined in a meaningful way. This means that real numbers, complex numbers, fuzzy numbers and even functions, operators, symbolic patterns or images can be used as cross-impacts if this makes sense for the problem under consideration. Nevertheless, the following section deals with the "normal" case of integer and real number cross-impacts, which is in practice the most important case.

## 3. Properties of CIB

From the mathematical point of view, the maximum impact score condition (the principle of consistency, cf. section 2) of CIB analysis establishes a new type of matrix calculus. It is necessary therefore to derive a set of basic definitions, basic properties, and basic laws which provide a framework for using this analysis tool. They are described and proved in this section.

**Definitions**

I) A CIB matrix of $N$ descriptors is a $N \times N$ - hypermatrix. The element $C_{ij}$ of the hypermatrix ("judgement section") is a $s_i \times s_j$ - matrix. $s_i$ is the number of states of descriptor $i$. $s_i$ are positive integers with may be different for the descriptors[1]. Together they build the "state-vector" $\underline{s}$. The entry in the cell $k,l$ of the judgement section $i,j$ - $C_{ij}(k,l)$ - is called a "cross-impact judgement". For the purposes of this paper, the cross-impact judgements are assumed to be real numbers or integer numbers. An extension is possible and discussed in section 2.

II) A CIB matrix with zero diagonal judgement sections is called a *regular CIB matrix*. Otherwise, the CIB matrix is an *extended CIB matrix*.

III) A row of a judgement section $\{C_{ij}(k,l), l = 1...s_j\}$ is called a *"judgement group"*. A judgement group with zero sum, i.e.

$$\sum_{l=1} C_{ij}(k,l) = 0 \qquad (2)$$

is called a *"standardized judgement group"*. If all judgement groups of a matrix are standardized, the matrix is called a *"standardized CIB matrix"* (Table 1 shows an example of a standardized CIB matrix).

IV) If

$$C_{ij}(k,l) = C_{ji}(l,k) \qquad (3)$$

is valid for all $i, j, k, l$ then $C$ is called a *"symmetric CIB matrix"*. If

$$C_{ij}(k,l) = -C_{ji}(l,k) \qquad (4)$$

is valid for all $i, j, k, l$ then $C$ is called an *"antisymmetric CIB matrix"*. If

$$C_{ij}(k,l) = C_{ij}(1 + s_i - k, 1 + s_j - l) \qquad (5)$$

is valid for all $k, l$ then $C_{ij}$ is called a *"polarized judgement section"* (cf. Table 2). If all judgement sections of a CIB matrix are polarized then the matrix is called a *"polarized CIB matrix"*. In Tab. 1, all judgement sections are polarized with the exception $C_{42}$. The matrix $C^+$ with

$$C_{ij}^+(k,l) = -C_{ji}(l,k) \quad (6)$$

is called the adjunct matrix of $C$. An antisymmetric CIB matrix is self-adjunct. Furthermore, Eq. 6 makes clear that $(C^+)^+ = C$. As usual, $C^T$ with

$$C_{ij}^T(k,l) = C_{ji}(l,k) \quad (7)$$

is called the transposed matrix of $C$.

---

[1] The case that all $s_i$ are equal to one is excluded.

**Table 2:** Examples of polarized judgement sections.

| k/l | 1 | 2 |   | k/l | 1 | 2 | 3 |   | k/l | 1 | 2 | 3 |
|---|---|---|---|---|---|---|---|---|---|---|---|---|
| 1 | +2 | -2 |   | 1 | -2 | +1 | +1 |   | 1 | +3 | -1 | -2 |
| 2 | -2 | +2 |   | 2 | +1 | +1 | -2 |   | 2 | 0 | 0 | 0 |
|   |   |   |   |   |   |   |   |   | 3 | -2 | -1 | +3 |

V) $z = [z_1, z_2, \ldots z_N]$ with integer $z_i$ and $1 \leq z_i \leq s_i$ is called a *"scenario"* of $C$. If $z$ satisfies all conditions (1) then $z$ is called a *"solution"* (*"consistent scenario"*) of $C$. The set of the $R = s_1 \cdot s_2 \cdot \ldots \cdot s_N$ possible scenarios of $C$, $\{z_1=1..s_1, z_2=1..s_2, \ldots z_N=1..s_N\}$ is called the *"set of combinatorial scenarios"*. A scenario $z$ highlighting the rows of $C$ (cf. Tab. 1) is abbreviated as $\langle z|$. A scenario $z$ highlighting the columns of $C$ is abbreviated by $|z\rangle$.

The sum

$$\theta_{jl} = \sum_i C_{ij}(z_i, l) \quad (8)$$

is called the *"impact score"* of state $l$ of descriptor $j$ with respect to scenario $z$. All impact scores of descriptor $j$ together make up the *"impact balance"* $\Theta_j$ of the descriptor $j$.

VI) The difference $\gamma_j$ between the impact score of the scenario state and the highest impact score of all other states of the descriptor $j$ is called the *"consistency"* of the descriptor $j$ ($\gamma_j \geq 0$ for a consistent scenario). The lowest consistency of all descriptors of a scenario, $\Gamma = \text{Min}\{\gamma_j\}$, is called the consistency of the scenario ($\Gamma \geq 0$ for a consistent scenario). A consistent scenario with $\Gamma = 0$ is also called a *"marginal consistent scenario"*. If $\gamma_j < 0$ then $\omega_j = -\gamma_j$ ($\omega_j = 0$ else) is also called the *"inconsistency"* of the descriptor $j$. If $\Gamma < 0$ then $\Omega = -\Gamma$ ($\Omega = 0$ else) is called the inconsistency of the scenario. I.e. in Table 1, the descriptor consistencies $\gamma_j$ are: -1, -2, +5, +2, +2. The scenario consistency $\Gamma$ is -2. The descriptor inconsistencies $\omega_j$ are: 1, 2, 0, 0, 0. The scenario inconsistency $\Omega$ is 2.

VII) The scalar product of two scenarios $\underline{v}$ and $\underline{z}$ is defined by (cf. Tab. 3)

$$\langle v | z \rangle = \sum_{ij}^{N} C_{ij}(v_i, z_j) \quad . \quad (9)$$

**Table 3:** The scalar product $\langle v | z \rangle$ of two scenarios $\underline{v} = [a_1, b_1, c_3]$ and $\underline{z} = [a_2, b_1, c_2]$ sums the intersections of the highlighted rows and columns. This equals the sum of the impact scores for the impacts of scenario $\underline{v}$ on scenario $\underline{z}$ (see row "Balances").

If necessary for unambiguity, the underlying matrix can be indicated by an index: $\langle v | z \rangle_C$. The definition of the CIB scalar product obviously implies:

$$\langle v | z \rangle_C = \langle z | v \rangle_{C^T} \quad (10)$$

$$\langle v | z \rangle_{C+D} = \langle v | z \rangle_C + \langle v | z \rangle_D \quad (11)$$

$$\langle v | z \rangle_{aC} = a \langle v | z \rangle_C \quad (12)$$

$$\langle v|z\rangle_C = -\langle z|v\rangle_{C^+} \quad . \qquad (13)$$

$C$ and $D$ are CIB matrices with the same state vector. $a$ is a scalar. Addition and scalar multiplication of CIB matrices are defined in the same way as for usual matrix calculus. The scalar product of a scenario $z$ with itself is called *"total impact score (TIS)"* of the scenario. It equals the sum of the impact scores of a scenario (e.g. in the scenario shown in Tab. 1, the TIS is 0-1+5+2+2 = +8).

VIII) The succession operator $S$ changes the states of a scenario according to a given CIB matrix $C$ and a given adjustment rule (e.g. global succession adjustment, cf. section 2). $|z'\rangle = S\,|z\rangle$ is called the successor of $z$. $z$ is a predecessor of $z'$. Every scenario may have no predecessor, one predecessor, or more than one predecessor.

IX) $|z\rangle$ is a member of an *"attractor"* of $C$ with respect to a given adjustment rule if $|z\rangle = S^n\,|z\rangle$ is valid for $n = p, 2p, 3p, \ldots$. The constant $p$ is called the *"period"* of the attractor. The attractor is defined by the set $\{S^k\,|z\rangle, k = 1,\ldots p\,\}$. An attractor with $p > 1$ is called a *"cyclic attractor (cycle)"*. An attractor with $p = 1$ is a consistent scenario. The set of all predecessors of all members of an attractor is called its *"basin of attraction"*. This also includes the members of the attractor themselves. The number of scenarios in the basin of attraction is called the *"weight"* of the attractor. The succession sequence of any scenario concludes in an attractor (because of the finite number of combinatorial scenarios and the unambiguity of the succession procedure). The sequence of successors prior to entering an attractor is called a *"transient"*.

X) CIB matrices $C$ and $C'$ are called *"equivalent"* if they possess the same consistent scenarios, transients, cycles, and attractor weights.

**Invariance operations**

XI) **IO-1: The addition of the same number in all judgement cells of a judgement group yields a CIB matrix $C'$ which is equivalent (cf. X) to the original matrix $C$. I.e.**

$$C'_{ij}(k,l) = C_{ij}(k,l) + a_{ijk} \qquad (14)$$

**with arbitrary $a_{ijk}$ (*"addition invariance"*). IO-1 can be used to standardize an arbitrary matrix (cf. III).**

**Proof**

The impact scores of a transformed matrix $C'$ with respect to an arbitrary scenario $z$ read (cf. Eq. 8):

$$\theta'_{jl} = \sum_i C'_{ij}(z_i, l) \qquad (15)$$

With (14) follows

$$\theta'_{jl} = \sum_i (C_{ij}(z_i, l) + a_{ij\,z_i})$$
$$= \theta_{jl} + \sum_i a_{ij\,z_i} \qquad (16)$$

The difference between $\theta'_{jl}$ and $\theta_{jl}$ does not depend on $l$. This means that the transformation $C \to C'$ shifts the whole impact balance of a descriptor. It follows that the state(s) of maximum impact score with respect to $C$ are also the state(s) of maximum impact score with respect to $C'$. Therefore the adjustments of descriptor states are the same in $C$ and $C'$ in the course of a succession. This means that the attractors, transients and attractor weights of $C$ and $C'$ are identical.

XII) **IO-2: The multiplication of all judgement cells in the judgement groups of a descriptor column with the same positive number yields a CIB matrix $C'$ which is equivalent (cf. X) to the original matrix $C$. I.e.**

$$C'_{ij}(k,l) = b_j\,C_{ij}(k,l) \qquad (17)$$

**with arbitrary $b_j > 0$ (*"local multiplication invariance"*).**

### Proof

In the case of Eq. 17, the impact scores of a transformed matrix $C'$ with respect to an arbitrary scenario $\underline{z}$ read (cf. Eq. 8):

$$\theta'_{jl} = \sum_i b_j\, C_{ij}(z_i, l) = b_j\, \theta_{jl}\, . \quad (18)$$

The factor $b_j$ does not depend on $l$. This means that the transformation $C \to C'$ multiplies the whole impact balance of a descriptor by the same positive number. It follows that the state(s) of maximum impact score with respect to $C$ are also the state(s) of maximum impact score with respect to $C'$. Therefore the adjustments of descriptor states are the same in $C$ and $C'$ in the course of a succession. This means that the attractors, transients and attractor weights of $C$ and $C'$ are identical.

**XIII) IO-3: The multiplication of all judgement cells of a CIB matrix with the same positive number yields a CIB matrix $C'$ which is equivalent (cf. X) to the original matrix $C$. I.e.**

$$C'_{ij}(k, l) = b\, C_{ij}(k, l) \quad (19)$$

**with arbitrary $b > 0$ ("global multiplication invariance").** IO-3 is an obvious consequence of IO-2. It is formulated as a separate statement because of its frequent use in CIB application practice.

**XIV) IO-4: The addition of a number in all judgement cells of a column within a judgement section and the subtraction of the same number in all judgement cells of the same column within another judgement section yields a CIB matrix $C'$ which is equivalent (cf. X) to the original matrix $C$. I.e.**

$$C'_{ij}(k, l) = C_{ij}(k, l) + d_{ijl} \quad (20)$$

$d_{ijl}$ **satisfies the conditions $\sum_l d_{ijl} = 0$ and is arbitrary otherwise ("transfer invariance").**

### Proof

In the case of Eq. 20, the impact scores of a transformed matrix C' with respect to an arbitrary scenario $\underline{z}$ read (cf. Eq. 8):

$$\theta'_{jl} = \sum_i \left(C_{ij}(z_i, l) + d_{ijl}\right) = \theta_{jl} + \sum_i d_{ijl} \quad (21)$$

The sum on the right side is zero. This means that the transformation $C \to C'$ has no influence on the impact balances. It follows that the state(s) of maximum impact score with respect to $C$ are also the state(s) of maximum impact score with respect to $C'$. Therefore the adjustments of descriptor states are the same in $C$ and $C'$ in the course of a succession. This means that the attractors, transients and attractor weights of $C$ and $C'$ are identical.

**Other properties**

**XV) If $\underline{z}$ is a solution of $C$ then**

$$\langle z|z \rangle \geq \langle z|v \rangle \quad (22)$$

**is valid for every scenario $\underline{v}$, and vice versa: if Eq. 22 is valid for every scenario $\underline{v}$ then $\underline{z}$ is a solution of $C$. Inequality (22) is thus an alternative formulation of the principle of consistency (1). Global succession adjustment can be formulated in a similar way: If $\underline{z}'$ is a successor of $\underline{z}$ then**

$$\langle z|z' \rangle \geq \langle z|v \rangle \quad (23)$$

**is valid for every scenario $\underline{v}$.**

### Proof

Inequality (1) is valid for an arbitrary choice of $l$ if $\underline{z}$ is a consistent scenario, which we will denote $\underline{z}^c$ for clarity. We choose $l = v_j$ (the component $j$ of an arbitrary scenario $\underline{v}$):

$$\sum_i C_{ij}(z_i^c, z_j^c) \geq \sum_i C_{ij}(z_i^c, v_j) \quad (24)$$

and sum (24) over all $j$. This yields

$$\sum_{ij} C_{ij}(z_i^c, z_j^c) \geq \sum_{ij} C_{ij}(z_i^c, v_j) \quad (25)$$

and with the definition of the scalar product (9):

$$\langle z^c | z^c \rangle \geq \langle z^c | v \rangle . \quad (26)$$

So far it has been proved that every consistent scenario $\underline{z}^c$ satisfies (22). To make (22) a full equivalent of the consistency principle (1), it also has to be proved that no inconsistent scenario $\underline{z}^{ic}$ satisfies (22). This can be done by showing that for each inconsistent scenario $\underline{z}^{ic}$ at least one scenario $\underline{v}$ exists for which (22) is untrue. This is the case for the successor of $\underline{z}^{ic}$ which we will denote $\underline{w}$. The adjustment rules of CIB succession (cf. section 2) imply that $\theta_{jz_j^{ic}} = \theta_{jw_j}$, i.e.

$$\sum_i C_{ij}(z_i^{ic}, z_j^{ic}) = \sum_i C_{ij}(z_i^{ic}, w_j) \quad (27)$$

if $j$ is a descriptor that is not adjusted ($w_j = z^{ic}_j$). If descriptor $j$ is adjusted (and this is the case for at least one descriptor because $\underline{z}^{ic}$ is an inconsistent scenario) then the impact score is higher after the adjustment and

$$\sum_i C_{ij}(z_i^{ic}, z_j^{ic}) < \sum_i C_{ij}(z_i^{ic}, w_j) \quad (28)$$

is true. Summing up all $j$ yields

$$\sum_{ij} C_{ij}(z_i^{ic}, z_j^{ic}) < \sum_{ij} C_{ij}(z_i^{ic}, w_j) \quad (29)$$

or, shorter:

$$\langle z^{ic} | z^{ic} \rangle < \langle z^{ic} | w \rangle . \quad (30)$$

This is in contradiction with (22), showing that (22) is true for all consistent scenarios, but not for an inconsistent scenario. Thus (22) is a full equivalent to the consistency principle (1).

XVI) **Every CIB matrix has at least one attractor (consistent scenario and/or cycle).** This is a direct consequence of the finite number of system states $R = s_1 \cdot s_2 \cdot \ldots \cdot s_N$.

XVII) **The weight of an attractor is greater than or equal to the period of the attractor.** This is an obvious consequence of the definition of the weight (cf. IX).

XVIII) **The TIS of a consistent scenario of a standardized CIB matrix is always non-negative** (cf. III and VII).

**Proof**

The impact balances of a standardized CIB matrix sum to zero for each descriptor, because they are calculated by the addition of zero-sum judgement sections. I.e. with (2) and (8):

$$\sum_l \theta_{jl} = \sum_{il} C_{ij}(z_i, l)$$
$$= \sum_i \underbrace{\sum_l C_{ij}(z_i, l)}_{\substack{=0 \ if \ C \\ standardized}} = 0$$

(31)

The impact scores of the states of a consistent scenario must be zero or positive under such circumstances because the states of a consistent scenario are the states of maximum impact score for each descriptor, and the maximum of a set of numbers with zero-sum cannot be negative. Therefore, the total impact score of a consistent scenario of a standardized CIB matrix is a sum of non-negative numbers, and so it is non-negative, too. This statement is also an implication of Eq. 39.

XIX) **If $\underline{z}$ is a solution of $C$ and $\underline{z}^+$ is a solution of $C^+$ then the TIS of $\underline{z}$ is greater than or equal to the TIS of $\underline{z}^+$ with regard to $C$ (cf. IV and VII). I.e.**

$$\langle z | z \rangle_C \geq \langle z^+ | z^+ \rangle_C . \quad (32)$$

**Proof**

We assume that $\underline{z}$ is a solution of $C$ and $\underline{z}^+$ is a solution of $C^+$. Then

$$\langle z | z \rangle_C \geq \langle z | v \rangle_C \quad (33)$$

and

$$\langle \underline{z}^+|\underline{z}^+\rangle_{C^+} \geq \langle \underline{z}^+|\underline{v}'\rangle_{C^+} \quad (34)$$

is valid for arbitrary scenarios $\underline{v}$ and $\underline{v}'$ because of (22). Now we choose $\underline{v} = \underline{z}^+$ and $\underline{v}' = \underline{z}$:

$$\langle \underline{z}|\underline{z}\rangle_C \geq \langle \underline{z}|\underline{z}^+\rangle_C \quad (35)$$

$$\langle \underline{z}^+|\underline{z}^+\rangle_{C^+} \geq \langle \underline{z}^+|\underline{z}\rangle_{C^+} \quad (36)$$

In (36) we use (13) and we obtain (after multiplying the inequality by $-1$):

$$\langle \underline{z}^+|\underline{z}^+\rangle_C \leq \langle \underline{z}|\underline{z}^+\rangle_C \quad (37)$$

Together with (35) we finally obtain:

$$\langle \underline{z}|\underline{z}\rangle_C \geq \langle \underline{z}^+|\underline{z}^+\rangle_C \quad (38)$$

**XX)** **In standardized CIB matrices with state vector $\underline{s}$ exists a lower bound of the TIS (cf. VII):**

$$\langle \underline{z}|\underline{z}\rangle \geq \Gamma \sum_i \frac{s_i - 1}{s_i}, \quad (39)$$

**valid for every scenario $\underline{z}$ (consistent or inconsistent). $\Gamma$ is the consistency of $\underline{z}$ (cf. VI).**

**Proof**

We consider the impact balance of descriptor $j$ with $s_j$ states, in which the state of maximum impact score has the impact score $\alpha_j$ and all other states have the impact scores $\varepsilon_1, \varepsilon_2, \ldots \varepsilon_{Sj-1}$. $\varepsilon_{max}$ is the maximum value of all $\varepsilon$. The consistency $\gamma_j$ of this descriptor is (cf. VI):

$$\gamma_j = \alpha_j - \varepsilon_{max} \quad (40)$$

If all judgement groups are standardized, the impact balance is standardized too (because it is the sum of standardized judgement groups, cf. Eq. 31), and

$$\alpha_j + \varepsilon_1 + \varepsilon_2 + \ldots \varepsilon_{Sj-1} = 0 \quad (41)$$

is valid. If we make use of $\varepsilon_{max}$, we can write:

$$\alpha_j + (s_j-1)\, \varepsilon_{max} \geq 0 \quad (42)$$

and, with (40),

$$s_j\, \alpha_j \geq (s_j-1)\, \gamma_j \quad (43)$$

and

$$\alpha_j \geq \frac{s_j - 1}{s_j} \gamma_j. \quad (44)$$

$\alpha_j$ is the contribution of descriptor $j$ to the TIS of the scenario (cf. VII). Therefore (cf. VI):

$$TIS = \sum_j \alpha_j \geq \sum_j \frac{s_j - 1}{s_j} \gamma_j$$

$$\geq \sum_j \frac{s_j - 1}{s_j} Min(\underline{\gamma}) = \Gamma \sum_j \frac{s_j - 1}{s_j}. \quad (45)$$

**XXI)** **If $\underline{z} = [z_1, z_2, \ldots z_N]$ is a solution of a polarized CIB matrix $C$ (cf. IV, Eq. 5) then the inverse scenario $\underline{z}^I = [1+s_1-z_1, 1+s_2-z_2, \ldots 1+s_N-z_N]$ is also a solution of $C$. As a consequence, a polarized CIB matrix with an even state number for at least one descriptor always has an even number of solutions.**

**Proof**

We check the consistency of the inverse scenario $\underline{z}^I = [1+s_1-z_1, 1+s_2-z_2, \ldots 1+s_N-z_N]$. If $\underline{z}^I$ is consistent,

$$\sum_i C_{ij}(1 + s_i - z_i, 1 + s_j - z_j)$$

$$\geq \sum_i C_{ij}(1 + s_i - z_i, l) \quad (46)$$

must be true for all $j, l$ (cf. (1)). Because $C$ is assumed to be a polarized CIB matrix, Eq. 5 is valid and can be used on both sides of (46). The result is:

$$\sum_i C_{ij}(z_i, z_j) \geq \sum_i C_{ij}(z_i, l') \quad (47)$$

valid for all $j, l'$ ($l' = 1 + s_j - l$), which is the condition that $\underline{z}$ is a consistent scenario. That means that $\underline{z}^I$ is a consistent scenario if $\underline{z}$ is consistent.

Therefore, the solutions of a polarized CIB matrix can be arranged in pairs and the total number of consistent scenarios must be an even number. Exception: if all $s_i$ are odd numbers, consistent scenarios with $\underline{z} = \underline{z}^I$ can occur (e.g. $\underline{z} = [2,2,2..2]$ in a $s_i=3$-matrix). In this case the number of consistent scenarios may also be odd for polarized CIB matrices.

**XXII)** For a regular symmetric CIB matrix $C$ (cf. II, IV), the scenario with the highest total impact score of all combinatorial scenarios is always a consistent scenario. This means also that a regular symmetric CIB matrix has always at least one consistent scenario. Further, all consistent scenarios of a regular symmetric CIB matrix are local maxima of the TIS.

**Proof**

In order to prove the consistency of the maximum TIS scenario for a regular symmetric CIB matrix we consider the impact of the local succession adjustment (cf. section 2) on the TIS of an inconsistent scenario $z$ (with inconsistency $\Omega$, cf. VI) for this matrix type. For the purpose of this proof we introduce the convention that only one descriptor is adjusted if there are more than one descriptor which share the maximum inconsistency ("single adjustment rule"). Let m be the descriptor of highest inconsistency $(\omega_m = \Omega)$ in which the adjustment takes place. The TIS of the scenario before the adjustment is applied may be written as:

$$\langle z|z \rangle = \sum_{ij} C_{ij}(z_i, z_j) = \sum_{\substack{i \neq m \\ j \neq m}} C_{ij}(z_i, z_j) + \sum_{i \neq m} C_{im}(z_i, z_m) + \sum_{j \neq m} C_{mj}(z_m, z_j) \quad . \quad (48)$$

Here we made use of $C_{mm} = 0$ (regular CIB matrix). The two last sums are equal because $C$ is assumed to be a symmetric CIB matrix. It follows:

$$\langle z|z \rangle = \sum_{\substack{i \neq m \\ j \neq m}} C_{ij}(z_i, z_j) + 2 \sum_{i \neq m} C_{im}(z_i, z_m) \quad (49)$$

The last sum is the impact score of state $z_m$ (cf. Eq. 8). Now we regard the TIS of $z'$ (the successor of $z$ with respect to the used adjustment rule). $z'$ has the same states as $z$ with the exception of descriptor $m$, whose state is $z'_m$ instead of $z_m$:

$$\langle z'|z' \rangle = \sum_{\substack{i \neq m \\ j \neq m}} C_{ij}(z'_i, z'_j) + 2 \sum_{i \neq m} C_{im}(z'_i, z'_m)$$

$$= \sum_{\substack{i \neq m \\ j \neq m}} C_{ij}(z_i, z_j) + 2 \sum_{i \neq m} C_{im}(z_i, z'_m) \quad (50)$$

The first first sum of (50) is the same as in (49) because this sum does not include the index m (the only descriptor which distinguishes $z$ and $z'$). The second sum is the impact score of the new state $z'_m$ which is selected by the adjustment rule. Because the descriptor m was assumed to show the inconsistency $\Omega > 0$, we know the exact difference between both impact scores (cf. VI):

$$\sum_{i \neq m} C_{im}(z_i, z'_m) = \sum_{i \neq m} C_{im}(z_i, z_m) + \Omega . \quad (51)$$

Eq, (49)-(51) lead us to:

$$\langle z'|z' \rangle = \langle z|z \rangle + 2\Omega \quad . \quad (52)$$

(52) expresses that the successor of an inconsistent scenario has a higher TIS than its predecessor (in terms of the used adjustment rule and a regular symmetric CIB matrix). This means that –TIS is a Lapjunov function[2]. (52) implies further that a regular symmetric CIB matrix cannot possess cyclic attractors with respect to the "single adjustment rule" because the recurrence of an inconsistent scenario in the course of the succession is impossible if the TIS increases monotonously.

The scenario with the highest TIS cannot increase its TIS further and hence must be a consistent scenario. This implies that every regular symmetric CIB matrix must have at least one solution (a consistent scenario) because it is inevitable that one of the combina-torial scenarios has the highest TIS of all consistent scenarios. If more than one scenario possesses the highest TIS simultaneously, all of them must be consistent scenarios.

---

[2] A function $L$ is called a Lapjunov function if $\Delta L \leq 0$ and $L \geq L_{min}$ is valid in the course of a dynamic process [13]. The existence of a Lapjunov function ensures that a system possesses a global stable minimum (as a practical consequence, regular symmetric CIB matrices can be used to describe systems which minimize or maximize a certain variable).

Eq. 51 shows also that the change of a single descriptor state of a consistent scenario ($\Omega=0$) will never increase the TIS in case of a regular symmetric CIB matrix. I.e. the consistent scenarios of a regular symmetric CIB matrix are local maxima of the TIS.

XXIII) **An antisymmetric CIB matrix $C$ (cf. IV) possesses at most one consistent scenario with consistency $\Gamma > 0$ (cf. VI). Furthermore, the TIS is zero for all combinatorial scenarios of an antisymmetric CIB matrix.**

**Proof**

If the equality is valid in (1) for one or more combinations of the values of $j$ and $l$, then the solution is a marginal consistent scenario ($\Gamma = 0$, cf. VI). Vice versa this means that the condition of consistency is stronger for a consistent scenario $\underline{z}$ with $\Gamma > 0$:

$$\sum_i C_{ij}(z_i, z_j) > \sum_i C_{ij}(z_i, l) \quad (53)$$

(valid for all $j, l$) and the alternative formulation of the principle of consistency is:

$$\langle z|z \rangle > \langle z|v \rangle \quad . \quad (54)$$

In this case, the relation (32) between the TIS of the solution of the matrix and the TIS of the solution of the adjunct matrix is stronger because it is restricted to solutions with $\Gamma > 0$:

$$\langle z|z \rangle_C > \langle z^+|z^+ \rangle_C \quad . \quad (55)$$

(54) and (55) can be proved by repeating the proof of (22) and repeating the proof of (32), using (53) instead of (1).

Now we assume that $C$ is antisymmetric and possesses more than one solution with consistency $\Gamma > 0$, and $\underline{z}_a$ and $\underline{z}_b$ are two of them. Because an antisymmetric CIB matrix $C$ is self-adjunct ($C = C^+$, cf. IV), $\underline{z}_a$ and $\underline{z}_b$ are also solutions of $C^+$. Using $\underline{z} = \underline{z}_a$ and $\underline{z}^+ = \underline{z}_b$ in (55) yields

$$\langle z_a|z_a \rangle_C > \langle z_b|z_b \rangle_C \quad (56)$$

whereas using $\underline{z} = \underline{z}_b$ and $\underline{z}^+ = \underline{z}_a$ in (55) yields

$$\langle z_b|z_b \rangle_C > \langle z_a|z_a \rangle_C \quad . \quad (57)$$

(56) and (57) is a contradiction. This proves that an antisymmetric CIB matrix cannot possess more than one solution with $\Gamma > 0$[3].

It is nearly obvious that TIS = 0 must be true for all combinatorial scenarios in the case of an antisymmetric CIB matrix. To derive this statement in a formal way, we chose $\underline{v} = \underline{z}$ in Eq. 13 and we obtain:

$$\langle z|z \rangle_C = -\langle z|z \rangle_{C^+} \quad (58)$$

valid in general, i.e. for every CIB matrix. In the case of an antisymmetric (self-adjunct) CIB matrix, $C = C^+$ is true and (58) leads us to:

$$\langle z|z \rangle_C = -\langle z|z \rangle_C \quad \Rightarrow \quad \langle z|z \rangle_C = 0 \quad . \quad (59)$$

XXIV) **For each qualitative (i.e. finite state) analysis task, whose solution can be described in a meaningful way by finite state scenarios, cycles or transients, at least one CIB matrix exist which provides exactly the correct consistent scenarios, cycles and transients ("CIB theorem of universality"). This means that the CIB framework is suitable as a universal tool in qualitative systems analysis, although it was formally developed as an analysis tool for pair interaction systems with additive effects.**

**Proof**

To prove this we show that CIB is equivalent to a Turing machine, or universal computer. To this aim it suffices to show that an algorithm can construct two different Boolean functions, including negation (cf. [17], p. 38). We show in Tab. 4 that CIB can execute the Boolean functions AND and NOT. Mapping the set of combinatorial scenarios to their re-spective successors can be formulated by a set of Boolean functions.

---

[3] In the generic case, an antisymmetric CIB matrix has no consistent scenario, but exceptions within the limited range of XXIII exist.

**Tab. 4:** The Boolean functions $x_2$ = NOT $x_1$ and $x_3$ = $x_1$ AND $x_2$, constructed by CIB matrices. The list of solutions of these CIB matrices is easy to construct by the method demonstrated in Tab. 1 and corresponds to the value table of the respective Boolean function.

| NOT | $x_1$ | | $x_2$ | |
|---|---|---|---|---|
| | 0 | 1 | 0 | 1 |
| $x_1$ 0 | | | -1 | 1 |
| 1 | | | 1 | -1 |
| $x_2$ 0 | 0 | 0 | | |
| 1 | 0 | 0 | | |

Consistent scenarios:

|    | $X_1$ | $X_2$ |
|----|-------|-------|
| 1) | 0     | 1     |
| 2) | 1     | 0     |

| AND | $x_1$ | | $x_2$ | | $x_3$ | |
|---|---|---|---|---|---|---|
| | 0 | 1 | 0 | 1 | 0 | 1 |
| $x_1$ 0 | | | 0 | 0 | 2 | -2 |
| 1 | | | 0 | 0 | -1 | 1 |
| $x_2$ 0 | 0 | 0 | | | 2 | -2 |
| 1 | 0 | 0 | | | -1 | 1 |
| $x_3$ 0 | 0 | 0 | 0 | 0 | | |
| 1 | 0 | 0 | 0 | 0 | | |

Consistent scenarios:

|    | $X_1$ | $X_2$ | $X_3$ |
|----|-------|-------|-------|
| 1) | 0     | 0     | 0     |
| 2) | 0     | 1     | 0     |
| 3) | 1     | 0     | 0     |
| 4) | 1     | 1     | 1     |

It follows that at least one CIB matrix exists for every requested mapping which executes this mapping. As a consequence, for every given set of solutions, cycles, weights and transients, at least one CIB matrix exists which leads to exactly this set when it is evaluated. It should be stressed that there is no guarantee that an analysis task with $N$ descriptors and $s$ states can be represented by a CIB matrix with $N$ descriptors and $s$ states. The representation of complex Boolean functions by CIB may require the use of auxiliary descriptors and auxiliary states. This may lead to an increase in the dimension of the matrix.

## 4. Summary

In this paper we analysed the mathematical background of the Cross-Impact Balance Analysis (CIB Analysis), a method for developing future scenarios by analysing qualitative expert statements on the interdependences of multidisciplinary systems. The CIB principle of consistency defines a new type of matrix calculus. The basics of an appropriate formalism have been developed which enable the analysis of the mathematical properties of this matrix type. Among the results is

- a set of invariance operations. They help understand, which types of judgement changes are able to influence the analysis results and which don't;

- a set of rules concerning symmetric CIB matrices. Among others they demonstrate that a CIB matrix can operate as a optimizing system;

- a universality theorem expressing that every finite state systems analysis can in principle be conducted in an exact way by CIB. The linear superposition of impacts used in CIB's impact balances doesn't cause a restriction of generality in a finite-state system.

The list of mathematical properties of CIB described in section 3 is not expected to be exhaustive and it definitely requires expansion. Because of the usefulness of the CIB concept in multidisciplinary systems analysis this work could be worth doing in the future.